\documentstyle[12pt]{article}
\begin{document}
\baselineskip=15pt
\begin{center}
{\Large {\bf The Japanese approach to the Shimura - Taniyama conjecture}}\\
\vspace{0.5cm}
{\bf H. Gopalkrishna Gadiyar$^1$, R. Padma$^1$ and H.S. Sharatchandra$^2$}\\
\end{center}
\vspace{1cm}
\begin{flushleft}
$^1$ AU-KBC Research Centre,
M.I.T. Campus, Anna University, \\
~~~Chromepet, Chennai 600 044 INDIA\\
~~~E-mail:gadiyar@au-kbc.org, padma@au-kbc.org\\
~~\\
$^2$ Institute of Mathematical Sciences, C.I.T. Campus,Taramani P.O., \\
~~~Chennai 600 113 INDIA\\
~~~E-mail: sharat@imsc.res.in
\end{flushleft}

\begin{center}
{\bf Abstract}
\end{center}

In this note we point out links between the Shimura - Taniyama conjecture and certain ideas in physics. Since all the seminal references are by strange coincidence Japanese we wish to call this the Japanese approach. The note elaborates on some inspired comments made by Barry Mazur in his popular article ``Number theory as gadfly."

\newpage

In his article ``Number theory as gadfly" [1], Barry Mazur makes certain perceptive comments: {\it As I shall not have the time to make clear, but hope, atleast to make believable, the conjecture of Shimura - Taniyama - Weil is a profoundly unifying conjecture - its very statement hints that we may have to look to diverse mathematical fields for insights or tools that may lead to its resolution. 

It does not seem unnatural to look to differential geometry for progress with this conjecture, or to partial differential equations and the study of the eigen value problem for elliptic operators, or to the representation theory of reductive groups.... It would be no surprise if ideas from the classical theory of one complex variable and the Mellin transform were relevant, or of Algebraic Geometry .....But perhaps one should also look in the direction of Kac - Moody algebras, loop groups or $\cal{D}$ - modules, perhaps to ideas that have been, or will be, imported from Physics...... }

In this note we point out that such a connection can be easily made. The comments quoted above can be sharpened to the use of the techniques of formal groups, conformal field theory and the theory of nonlinear differential equations.

We view the essence of the Shimura - Taniyama conjecture as extracting arithmetical information from a geometric object defined analytically. An elliptic curve is essentially defined given a function $\wp (z)$. As this function obeys a nonlinear differential equation it is evident that the theory of nonlinear differential equations would be an essential tool. 

The conjecture of Shimura - Taniyama says that there is another parametrization which contains arithmetical information. This is given by  the mapping [2]
$$
\alpha (z) = \wp (\sum_{n=1}^\infty \frac{a(n)}{n} e^{2\pi inz}) 
$$
$$
\beta (z) = \wp ^{\prime}(\sum_{n=1}^\infty \frac{a(n)}{n} e^{2\pi inz})
$$
The $a(n)$ are the objects encoding arithmetical information. 

We wish to point out that the theory of the KP - hierarchy provides the key to linking the Shimura - Taniyama parametrization to the theory of soliton equations. In the KP - hierarchy the essential object of study is the so called $\tau $ function which is essentially $\theta (\sum_{k=1}^\infty a_k t_k )$. This has already been used to settle the Schottky problem by Mulase [3] and Shiota [4]. 

The next ingredient is the theory of formal groups. Honda [5, 6] showed the isomorphism between the formal group arising from the analytic parametrization and the formal group arising from the arithmetical parametrization. In fact, he has even provided explicit constructions. Finally the connection between formal groups and the $\tau $ function has been made in the paper [7]. In this paper Hecke like operators are constructed. It essentially involves mappings of the $t_k$. One can construct the strict analogue of Hecke operators also. 

Hence the observations of Barry Mazur can be made precise. Essentially one goes from $\wp (z)$ to the $\tau$ function which contains arithmetical information. 

It is well known that the $\tau$ function is linked to a rich variety of mathematical objects like conformal field theory, infinite dimensional Lie groups, vertex operators, and so on fulfilling Barry Mazur 's predictions. The references given are those that influenced the authors and are not meant to be exhaustive.

We have written this brief version to enable the community of mathematical physicists to develop what seems a rich mine. 

GHG wishes to thank Professor N. D. Haridass for introducing him to the work of Miwa, Jimbo and Date [8]. 

\vspace{0.5cm}
\noindent{\bf References}
\begin{enumerate}
\item[1] B. Mazur, Number theory as gadfly, Amer. Math. Monthly, {\bf 98}(1991), 593-610
\item[2] D. Goldfeld, Modular elliptic curves and Diophantine problems, Number Theory, Proceedings of the First Conference of the Canadian Number theory Association held at the Banff Centre, Banff, Alberta, April 17-27, 1988, R. Mollin (Ed.), Walter de Gruyter Inc., 157-175
\item[3] M. Mulase, Cohomological soliton equations and Jacobian varieties, J. Diff. Geom., {\bf 19} (1984), 403-430
\item[4] T. Shiota, Characterization of Jacobian varieties in terms of soliton equations, Invent. Math. {\bf 83}(1986), 333-382
\item[5] T. Honda, Formal groups and zeta functions, Osaka J. Math. {\bf 5} (1968), 199-213
\item[6] T. Honda, Invariant differentials and L-functions. Reciprocity law for quadratic fields and elliptic curves over {\bf Q}, Rend. Sem. Math. Univ. Padova {\bf 49} (1973), 323-335
\item[7] T. Katsura, Y. Shimizu, amd K. Ueno, Formal groups and conformal field theory over Z, Advanced Studies in Pure Mathematics 19, 1989, Integrable Systems in Quantum Field Theory and Statistical Mechanics, Kinokuniya Shoten and Academic Press, 347-366
\item[8] T. Miwa, M. Jimbo and E. Date, Solitons: Differential equations, symmetries and infinite dimensional algebras, Cambridge University Press, 2000
\end{enumerate}
\end{document}